\documentclass[journal]{IEEEtran}

\usepackage{cite}
\usepackage{amsmath,amssymb,amsfonts}
\usepackage{bm}
\usepackage{graphicx}
\usepackage{textcomp}
\usepackage{mathtools}
\usepackage{siunitx}
\usepackage{booktabs}
\usepackage{multirow}
\usepackage{url}
\usepackage{xcolor}
\usepackage{balance}

\usepackage{algorithm}
\usepackage{algorithmic} 

\usepackage[caption=false]{subfig}
\usepackage{cuted}   
\usepackage{amsmath,amssymb}
\usepackage{amsmath}
\usepackage{graphicx} 
\usepackage{graphicx}
\usepackage{epstopdf} 
\usepackage{flushend}
\begin{document}

\title{\mbox{Rotatable Antenna Assisted Mobile Edge Computing}}

\author{Ji~Wang,~\IEEEmembership{Senior Member,~IEEE,}
        Hao~Chen,
        Yixuan~Li,
        Jun~Zhang,
        Xingwang~Li,~\IEEEmembership{Senior Member,~IEEE,}
        Ming~Zeng,
        and~Octavia~A.~Dobre,~\IEEEmembership{Fellow,~IEEE}%
\thanks{Ji Wang, Hao Chen, and Yixuan Li are with the Department of Electronics and Information Engineering, College of Physical Science and Technology, Central China Normal University, Wuhan 430079, China (e-mail: jiwang@ccnu.edu.cn; 1280904251ch@mails.ccnu.edu.cn; yixuanli@mails.ccnu.edu.cn).}%
\thanks{Jun Zhang is with the School of Physics and Electronic Engineering, Hubei University of Arts and Science, Xiangyang 441053, China, and also with the Hubei Provincial Engineering Research Center of Emergency Communication Technology and System, Xiangyang 441053, China (e-mail: hbuas\_zhangjun@hbuas.edu.cn). (Corresponding author: Jun Zhang.)}%
\thanks{Xingwang Li is with the School of Physics and Electronic Information Engineering, Henan Polytechnic University, Jiaozuo 454003, China (e-mail: lixingwang@hpu.edu.cn).}%
\thanks{Ming Zeng is with the Department of Electrical and Computer Engineering, Laval University, Quebec City, QC G1V 0A6, Canada (e-mail: ming.zeng@gel.ulaval.ca).}%
\thanks{Octavia A. Dobre is with the Faculty of Engineering and Applied Science, Memorial University, St. John's, NL A1C 5S7, Canada (e-mail: odobre@mun.ca).}%
}


\maketitle

\begin{abstract}
This paper investigates a rotatable antenna (RA) assisted mobile edge computing (MEC) network, where multiple users offload their computation tasks to an edge server equipped with an RA array under a time-division multiple access protocol. To maximize the weighted sum computation rate, we formulate a joint optimization problem over the RA rotation angles, time-slot allocation, transmit power, and local CPU frequencies. Due to the non-convex nature of the formulated problem, a \textbf{scenario-adaptive hybrid optimization algorithm} is proposed. Specifically, for the dynamic rotating scenario, where RAs can flexibly reorient within each time slot, we derive closed-form optimal antenna pointing vectors to enable a low-complexity sequential solution. In contrast, for the static rotating scenario where RAs maintain a unified orientation, we develop an alternating optimization framework, where the non-convex RA rotation constraints are handled using successive convex approximation iteratively with the resource allocation. Simulation results demonstrate that the proposed RA assisted MEC network significantly outperforms conventional fixed-antenna MEC networks. Owing to the additional spatial degrees of freedom introduced by mechanical rotation, the flexibility of RAs effectively mitigates the severe beam misalignment inherent in fixed-antenna systems, particularly under high antenna directivity.
\end{abstract}


\section{Introduction}
\IEEEPARstart{T}{he} rapid proliferation of Internet-of-Things (IoT) devices and artificial intelligence--driven applications is imposing unprecedented computational and latency requirements on sixth-generation wireless networks. The paradigm of mobile edge computing (MEC) has emerged as a pivotal architecture to mitigate these bottlenecks, primarily via the migration of computationally demanding workloads from capacity-limited terminals to proximal edge infrastructure~\cite{cui2025overview}. Nevertheless, the performance gains achievable by MEC are fundamentally limited by the quality of the wireless computation offloading links, making communication efficiency a critical bottleneck~\cite{wang2018joint}.

Most existing MEC systems rely on fixed antennas (FAs) at the base station. While directional antennas are desirable for their high gain, extended transmission range, and strong interference resistance, their inherent narrow beamwidths make them highly sensitive to directionality. Once a user's position deviates from the fixed pointing direction, the advantage of high gain instantly becomes a disadvantage, leading to a sharp decline in link quality. This physical rigidity often results in severe beam misalignment and degraded offloading efficiency, particularly in scenarios with uneven user distributions or dynamically changing channel conditions~\cite{7762913}.
\label{sec:system_model}
\begin{figure}[!t]
 \centering
 \includegraphics[width=0.7\columnwidth]{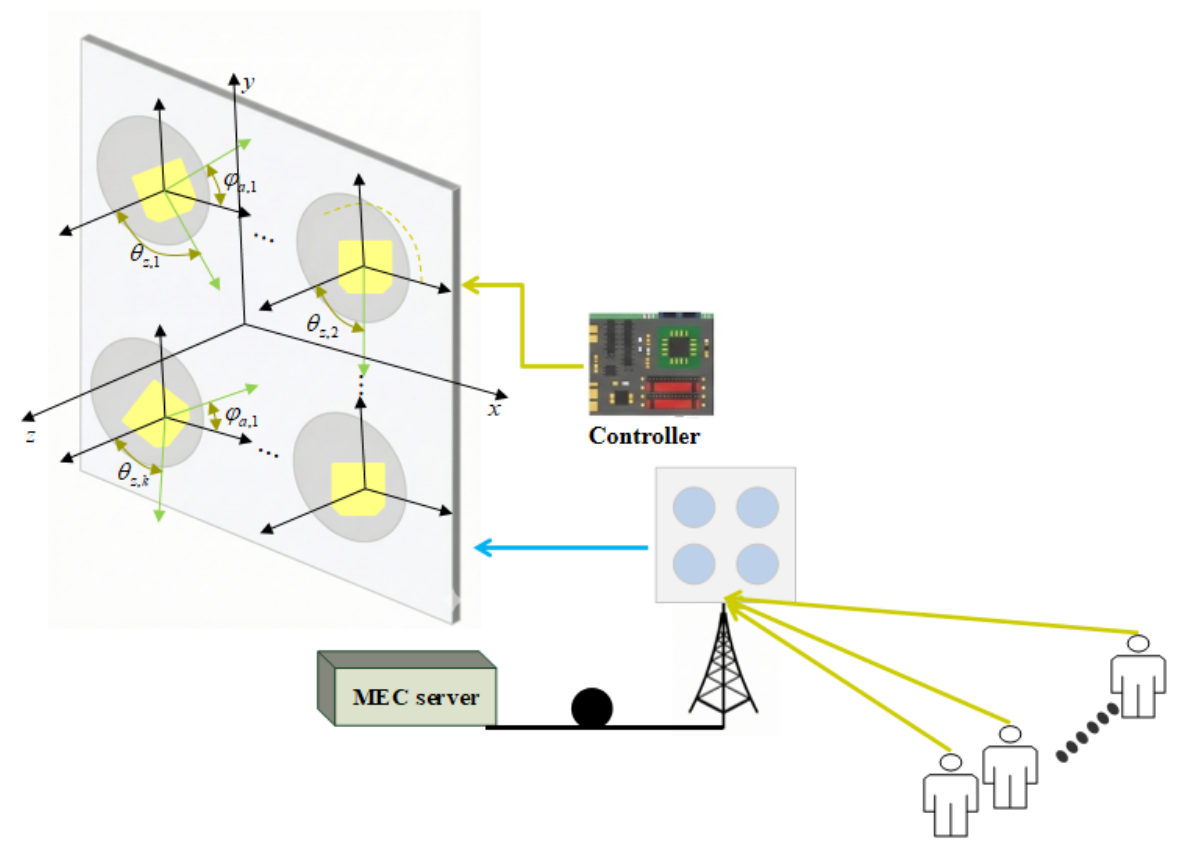}

\caption{The proposed RA assisted MEC system model.}
\label{fig:system_model}
\vspace{-5pt}
\end{figure}Although conventional multiple-input multiple-output architectures partially exploit spatial diversity, the fixed placement and orientation of antenna elements inherently constrain their ability to fully capitalize on spatial channel variations in data-intensive MEC applications.

To address these limitations, reconfigurable intelligent surfaces (RISs)~\cite{8741198,10702481,li2025multilayer,liang2025covert} and pinching antennas (PAs)~\cite{11106459,zeng2025resourceallocationpinchingantennasystems} explore spatial degrees of freedom, but they face challenges such as double-fading effects and hardware implementation complexity, respectively.
In contrast to RIS- and PA-based approaches that manipulate the propagation medium or radiation location, rotatable antennas (RAs) exploit spatial adaptability by mechanically steering the antenna boresight in three-dimensional (3D) space to align with dominant channel directions~\cite{ref1}. By leveraging mechanical degrees of freedom to trade for spatial degrees of freedom, RAs provide more precise and stronger beam alignment. This mechanism effectively mitigates the severe beam misalignment inherent in fixed-antenna systems, ensuring that the high gain of directional antennas is fully utilized even for dispersed users. Moreover, compared with movable antennas, which require continuous positional adjustments and incur significant mechanical and control complexity~\cite{ref2}, RAs offer a substantially lower-complexity solution by leveraging rotational motion only, making them particularly attractive for practical MEC deployments.

Recent studies have established a growing body of work on RA-assisted wireless systems, encompassing channel modeling~\cite{ref1}, efficient channel estimation strategies~\cite{ref3}, and advanced applications such as integrated sensing and communications~\cite{ref4}, physical-layer security~\cite{ref5}, and hybrid movable--rotatable antenna architectures~\cite{ref6}. Notably, these works have theoretically and experimentally verified that RA-assisted systems outperform conventional fixed-antenna systems in various scenarios~\cite{ref1,ref3,ref4,ref5}. It has been demonstrated that the performance gain of RAs increases with the antenna directivity factor, confirming that rotational flexibility is essential to unlock the potential of narrow-beam directional antennas. Furthermore, the feasibility of RA implementation has been validated through hardware prototypes and dynamic demonstrations~\cite{ref1}. However, the integration of RAs into MEC frameworks has received scant attention in the existing literature. In particular, existing MEC-oriented resource allocation frameworks do not account for the rotational degrees of freedom offered by RAs, nor do they investigate the synergistic orchestration of antenna orientation alongside transmission and processing resources to improve offloading efficiency.

 To fully unlock the potential of RAs in MEC, we formulate a joint resource optimization problem that maximizes the weighted sum computation rate by jointly optimizing the RA rotation angles, time-slot durations, transmit power allocation, and local CPU frequencies. To address the non-convex problem, we develop a \textbf{scenario-adaptive hybrid optimization algorithm} (SAHO). This strategy efficiently adapts the solution structure to the RA operating mode to balance performance and complexity. Specifically, for the \textit{dynamic rotating scenario}, we exploit the derived closed-form optimal pointing solution to decouple the problem, allowing for a fast sequential optimization of antenna orientation and offloading resources. Conversely, for the practical \textit{static rotating scenario}, where variables are intricately coupled, we propose an alternating optimization (AO) method. This method iteratively updates the RA rotation angles via successive convex approximation (SCA) and the offloading resources via convex optimization until convergence.

\vspace{-10pt}
\section{System Model}
Fig.~\ref{fig:system_model} depicts the considered uplink MEC framework, wherein a base station (BS) integrated with an edge server supports $M$ users, each having a single antenna. The BS utilizes a uniform planar array (UPA) composed of $K = K_x \times K_y$ RAs, with $K_x$ and $K_y$ representing the array dimensions along the x- and y-axes. Transmission is governed by a time-division multiple access (TDMA) protocol with a frame length of $T$.
\vspace{-10pt}

\subsection{Network Geometry and Channel Model}
The UPA is deployed in the x-y plane, with adjacent elements separated by a distance $\Delta$. The coordinate of the $k$-th RA is given by $\mathbf{w}_k = [k_x\Delta, k_y\Delta, 0]^T$, where $k_x$ and $k_y$ are indices relative to the array center. The $m$-th user is located at $\mathbf{u}_m = [r_m \sin\psi_m \cos\varphi_m, r_m \sin\psi_m \sin\varphi_m, r_m \cos\psi_m]^T$, where $r_m$ denotes the distance of user $m$ from the origin, and $\psi_m \in [0,\pi]$ and $\varphi_m \in \left[-\frac{\pi}{2},\,\frac{\pi}{2}\right]$ represent the zenith and azimuth angles of the $m$-th user relative to the coordinate origin, respectively.

Unlike fixed antenna systems, each RA $k$ can mechanically rotate its boresight direction, characterized by a unit pointing vector $\mathbf{f}_k \in \mathbb{R}^{3}$. Let $\theta_{z,k}$ and $\phi_{a,k}$ represent the zenith and azimuth definitions for the rotation of the $k$-th RA. Accordingly, its pointing vector is formulated as $f_k = [\sin \theta_{z,k} \cos \phi_{a,k}, \sin \theta_{z,k} \sin \phi_{a,k}, \cos \theta_{z,k}]^T$.
To avoid hardware coupling, the rotation is constrained by a maximum zenith angle $\theta_{\max}$, i.e., $0 \le \theta_{z,k} \le \theta_{\max}$. Here, $(\epsilon, \phi)$ denotes the incident angle tuple characterizing an arbitrary spatial direction relative to the RA's boresight axis. Due to the symmetry of the antenna pattern, the gain depends only on $\epsilon$. Accordingly, the effective antenna gain from user $m$ to RA $k$ follows $G_{k,m} = G_0 \cos^{2p}(\epsilon_{k,m})$, where $G_0$ is the maximum gain, $p$ is the directivity factor, and $\cos(\epsilon_{k,m}) \triangleq \mathbf{f}_k^T \mathbf{q}_{k,m}$ represents the cosine of the misalignment angle, with $\mathbf{q}_{k,m} \triangleq \frac{\mathbf{u}_m - \mathbf{w}_k}{\|\mathbf{u}_m - \mathbf{w}_k\|}$.

Assuming quasi-static flat fading, the overall multipath channel between user $m$ and the $k$-th RA at the BS is modeled as $h_{k,m}(\mathbf{f}_k)=\sqrt{L(d_{k,m})}\,G_{k,m}\,g_{k,m}$. Here, $L(d_{k,m})$ characterizes the large-scale attenuation, which is formulated as $L(d_{k,m}) = A_0(d_0/d_{k,m})^{\alpha_m}$. In this expression, $d_{k,m}$ represents the spatial separation between user $m$ and the $k$-th RA, while $A_0$ and $\alpha_m$ denote the channel gain at the reference distance $d_0 = 1$ m and the path-loss exponent, respectively.

The small-scale fading coefficient $g_{k,m}$ is modeled as a superposition of a deterministic line-of-sight (LoS) path and a scattered component, governed by the Rician factor $\kappa_m$. This is formulated as $g_{k,m} = \sqrt{\kappa_m/(\kappa_m+1)}\,\bar{g}_{k,m} + \sqrt{1/(\kappa_m+1)}\,\tilde{g}_{k,m}$, wherein $\overline{g}_{k,m} = e^{-j \frac{2\pi d_{k,m}}{\lambda}}$ signifies the LoS phase rotation, and $\tilde{g}_{k,m} \sim \mathcal{CN}(0, 1)$ captures the non-LoS Rayleigh fading effects. We define the pointing matrix collecting all RA boresight vectors as $\mathbf{F}\triangleq[\mathbf{f}_1,\mathbf{f}_2,\ldots,\mathbf{f}_K]$. Then, the channel vector from user $m$ to the BS is written as $\mathbf{h}_m(\mathbf{F}) \triangleq [h_{1,m}(\mathbf{f}_1), h_{2,m}(\mathbf{f}_2), \ldots, h_{K,m}(\mathbf{f}_K)]^T$. For analytical convenience, we assume that the BS has acquired the global channel state information of all links.
\vspace{-10pt}
\subsection{Computation and Offloading Models}
\label{sec:comp_model}

We adopt a partial offloading model where tasks are executed locally or offloaded to the RA assisted edge server via TDMA.

\subsubsection{Local Computing}
\label{ssec:local_comp}
For user $m$, let $f_m$ be the CPU frequency. We define the local CPU frequency vector for all users as $\mathbf{f} \triangleq [f_1, f_2, \dots, f_M]^T$. The locally processed data size is $R_{\text{loc},m} = T f_m / C$, and the energy consumption is $E_{\text{loc},m} = T r_c f_m^3$, with $T$ and $r_c$ representing the duration of the entire time frame and the effective capacitance parameter, respectively.

\subsubsection{Computation Offloading}
\label{ssec:offload_comp}
Under the TDMA scheme, users transmit in orthogonal slots $\tau_m$, eliminating inter-user interference. Considering a communication overhead factor $v_m > 1$, which accounts for channel coding redundancy and protocol headers, the offloaded data size $R_{off,m}$ is given by
\vspace{-5pt}
\begin{equation}
\small
R_{\text{off},m} = \frac{\tau_m B}{v_m} \log_2 \left( 1 + \frac{p_m \left\| \mathbf{h}_m(F) \right\|^2}{\sigma^2} \right),
\end{equation}
where $B$ is the bandwidth, $p_m$ is the transmit power, and $\sigma^2$ is the noise power. The corresponding energy consumption $E_{\text{off},m}$ is denoted as $E_{\text{off},m} = \tau_m (p_m + p_{c,m})$, where $p_{\text{c,m}}$ is the circuit power consumption.
\vspace{-10pt}
\subsection{Problem Formulation}
\label{ssec:problem_formulation}

We establish the computation rate maximization model (P1) by jointly optimizing the RA pointing matrix $\mathbf{F}$, local CPU frequency vector $\mathbf{f}$, time-slot allocation vector $\boldsymbol{\tau}$, and transmit power vector $\mathbf{p}$. Specifically, $\mathbf{F}$ determines the antenna pointing directions of the RAs subject to the physical rotation constraints, while $\mathbf{f}, \boldsymbol{\tau}, \text{and } \mathbf{p}$ coordinate the computation and communication resources under energy and latency budgets. The problem is formulated as
\begin{subequations}\label{eq:P1}
\begin{align}
\text{(P1)}: \quad
\max_{\mathbf{p}, \boldsymbol{\tau}, \mathbf{f}, \mathbf{F}} \quad
& \sum_{m=1}^{M} \big( R_{\text{loc},m} + R_{\text{off},m} \big) \label{eq:P1_obj} \\
\text{s.t.} \quad
& E_{\text{off},m} + E_{\text{loc},m} \le E_{\max}, \quad \forall m, \label{eq:P1_energy} \\
& \sum_{m=1}^{M} \tau_m \le T, \quad 0 \le \tau_m \le T, \quad \forall m, \label{eq:P1_time} \\
& p_m \ge 0, \quad f_m \ge 0, \quad \forall m, \label{eq:P1_pf} \\
& R_{\text{loc},m} + R_{\text{off},m} \ge R_{\min}, \quad \forall m, \label{eq:P1_min_bits} \\
& 0 \le \theta_{z,m} \le \theta_{\max}, \quad \forall m. \label{eq:P1_angle}
\end{align}
\end{subequations}

\vspace{-10pt}




\section{Joint Optimization Solution and Algorithm}
\label{sec:joint_opt}

The non-convexity inherent in (P1) arises principally from the complex interplay involving the RA pointing matrix $\mathbf{F}$, the resource allocation variables ($\{\mathbf{f}, \mathbf{p}, \bm{\tau}\}$), and the non-convex unit-modulus constraints on the pointing vectors. Considering the difficulty of directly solving the  problem P1, we decouple the problem into RA rotation optimization and offloading optimization subproblems. For RA rotation optimization, we separately address the dynamic and static rotatable antenna scenarios, accounting for the rotation speed of RAs relative to the TDMA frame length. In the offloading optimization stage, the pointing matrix is held fixed, and the remaining resource allocation problem becomes convex, allowing it to be solved efficiently.
\vspace{-10pt}

\subsection{RA Rotation Optimization}
\label{ssec:RA_opt}

In the RA rotation optimization procedure, we focus on optimizing the RA pointing matrix $\mathbf{F}$ to maximize the directional gain, while keeping the offloading parameters $\{\mathbf{f}, \mathbf{p}, \boldsymbol{\tau}\}$ fixed.
We consider two scenarios based on the RA's rotation timescale: 1) dynamic rotation, and 2) static rotation.

\subsubsection{Dynamic rotation}
Under the uplink rate within the time slot, the optimization of antenna orientation aims to maximize projection between $\mathbf{f}_k$ and $\mathbf{q}_{k,m}$, which can be formulated as
\begin{subequations}\label{eq:P2_full}
\begin{align}
\text{(P2)}: \quad
\max_{\mathbf{f}_k} \quad & \mathbf{f}_k^{T}\mathbf{q}_{k,m} \label{eq:P2_obj_sub} \\
\text{s.t.} \quad & \|\mathbf{f}_k\| = 1, \label{eq:P2_norm_sub} \\
& 0 \leq \arccos(\mathbf{f}_k^T \mathbf{e}_3) \leq \theta_{\text{max}}, \label{eq:P2_angle_sub}
\end{align}
\end{subequations}
where, $e_3 = [0, 0, 1]^T$ corresponds to the standard basis vector aligned with the z-axis.
The objective $\mathbf{f}_k^{T}\mathbf{q}_{k,m}=\cos(\epsilon_{k,m})$
represents the directional projection between the RA boresight and the user direction.
Maximizing this value is equivalent to minimizing the pointing deviation $\epsilon_{k,m}$. The (P2) is a constrained linear optimization over a unit sphere and admits a closed-form global optimum which is given by
\begin{equation}
\mathbf{f}_n^\star =
\begin{cases}
\mathbf{q}_{k,m}, &
    \text{if } \arccos(\mathbf{u}_{m,n}^{T}\mathbf{e}_3)\le \theta_{\max},\\[6pt]
\begin{bmatrix}
    \sin\theta_{\max}\cos\phi_{a,k}\\
    \sin\theta_{\max}\sin\phi_{a,k}\\
    \cos\theta_{\max}
\end{bmatrix}, &
    \text{otherwise},
\end{cases}
\label{eq:f_opt}
\end{equation}
where the rotation angles are calculated as
\begin{align}
    \theta_{z,k} &\triangleq \min \left\{ \arccos \left( \mathbf{q}_{k,m}^T \mathbf{e}_3 \right), \theta_{\max} \right\}, \label{eq:theta_calc} \\
    \phi_{a,k} &\triangleq \operatorname{arctan2} \left( \mathbf{q}_{k,m}^T \mathbf{e}_2, \mathbf{q}_{k,m}^T \mathbf{e}_1 \right), \label{eq:phi_calc}
\end{align}
and the standard basis vectors are defined as $\mathbf{e}_1=[1,0,0]^T$, $\mathbf{e}_2=[0,1,0]^T$, and $\mathbf{e}_3=[0,0,1]^T$.

When the user direction lies within the rotation range (i.e., $\arccos(\mathbf{q}_{k,m}^{T}\mathbf{e}_3)\le \theta_{\max}$), the RA boresight is steered to coincide exactly with the user's angular direction, thereby attaining the peak directional gain. Otherwise, the RA is rotated to its maximum zenith angle $\theta_{\max}$ while maintaining the azimuth alignment, achieving ``edge alignment'' within the feasible region.

Consequently, the optimal pointing matrix for all RAs in time slot $m$ is $\mathbf{F}^\star = [\mathbf{f}_1^\star, \mathbf{f}_2^\star, \ldots, \mathbf{f}_K^\star]$.
This result aligns with theoretical predictions, as the BS realizes the peak directional gain $KG_0$ provided that the boresight axis of every RA is strictly steered toward the user's angle of arrival, i.e., \( \mathbf{f}_k = \mathbf{q}_{k,m} \).

\subsubsection{\textit{Static rotation}}

However, the assumption of flexible re-orientation within each time slot is ideal but practically challenging. Due to the inherent mechanical inertia, the rotation speed of RAs is significantly slower than the time-slot switching frequency in TDMA protocols. Consequently, it is physically impossible for RAs to instantaneously realign their boresight for different users in consecutive slots. Instead, a unified antenna orientation must be determined to serve all users effectively throughout the entire timeframe. For a given offloading parameter $\{\mathbf{f}, \mathbf{p}, \boldsymbol{\tau}\}$, problem (P1) can be reformulated as
\vspace{-10pt}
\begin{subequations}\label{prob:P3}
\begin{align}
\text{(P3)} \quad \max_{\mathbf{F}} \quad
& \sum_{m=1}^M \left( R_{\text{loc},m} + R_{\text{off},m} \right) \label{eq:P3_obj} \\
\text{s.t.} \quad 
& 0 \le \theta_{z,k} \le \theta_{\max}, \quad \forall k. \label{eq:P3_const}
\end{align}
\end{subequations}

Crucially, the deflection angles primarily modulate the channel power gain via the projection term $\cos(\epsilon_{k,m})$. To facilitate the SCA-based algorithm, we rewrite the channel coefficient by grouping the constant terms and explicitly expressing the dependence on the pointing vector $f_k$. The channel is reformulated as
\begin{equation}
h_{m}(\mathbf{f}_k) = \tilde{\beta}_{m,k} \left( \mathbf{f}_k^{T} \mathbf{q}_{k,m} \right)^{p},
\label{eq:h_mk}
\end{equation}
where $\tilde{\beta}_{m,k} \triangleq \sqrt{L(d_{m,k})\,G_0}\,g_{k,m}$. 

By substituting the reformulated scalar channel elements from (8) into the system channel vector $\mathbf{h}_m(\mathbf{F})$, it becomes evident that the effective channel gain is directly determined by the RA pointing matrix $\mathbf{F}$. Consequently, the optimization in (P3) is equivalent to maximizing the offloading sum rate with respect to $\mathbf{F}$.
\vspace{-10pt}
\begin{subequations}\label{prob:P4}
\begin{align}
\text{(P4)} \quad \max_{\mathbf{F}} \quad 
& \sum_{m=1}^{M} \frac{\tau_m B}{v_m} \log_2 \left( 1 + \frac{p_m \| \mathbf{h}_m(\mathcal{F}) \|^2}{\tau_m \sigma^2} \right)
\label{eq:P4_obj} \\
\text{s.t.} \quad 
& \cos(\theta_{\max}) \le \mathbf{f}_k^T\mathbf{e}_3 \le 1, \quad \forall k, \label{eq:P4_angle} \\
& \|\mathbf{f}_k\| = 1, \quad \forall k. \label{eq:P4_norm}
\end{align}
\end{subequations}

Here, \eqref{eq:P4_angle} mirrors the constraint in \eqref{eq:P3_const}, while \eqref{eq:P4_norm} strictly enforces the unit-norm requirement on $f_k$. Since Problem (9) retains its non-convex character, rendering a direct solution intractable, we resort to the SCA framework to construct a convex surrogate, thereby iteratively approaching a local optimum. Specifically, during the $(i+1)$-th update, given the prior estimates $F^{(i)}$ and $R_{off}^{(i)}$, we linearize the logarithmic term in (9) via a first-order Taylor expansion around $f_k^{(i)}$, yielding the approximation $\Lambda_r^{(i+1)}(F)$ formulated below
\begin{equation}
\tilde{\mathbf{h}}_{m,k} = \frac{\partial h_{m,k} (\mathbf{f}_k^{(i)})}{\partial \mathbf{f}_k^{(i)}} = \tilde{\beta}_{m,k} p \left( (\mathbf{f}_k^{(i)})^T \mathbf{q}_{k,m} \right)^{p-1} \mathbf{q}_{k,m}.
\label{eq:gradient}
\end{equation}
Consequently, for the $(i+1)$-th iteration, problem (P4) is reformulated via approximation as
\begin{subequations}\label{prob:P5}
\begin{align}
\text{(P5)} \quad \max_{\mathbf{F}} \quad & \Lambda^{(i+1)}_{r}(\mathbf{F}) \label{eq:P5_obj} \\
\text{s.t.} \quad & \eqref{eq:P4_angle}, \eqref{eq:P4_norm}. \label{eq:P5_const}
\end{align}
\end{subequations}

\begin{figure*}[!t]
    \footnotesize
    \vspace{-5pt} 
    \begin{equation}
    \resizebox{0.85\textwidth}{!}{$
        \displaystyle
        \Lambda_r^{(i+1)}(\mathbf{F}) \triangleq
        \log_2 \!\left( 1 + \frac{p_m \|\mathbf{h}_m(\mathbf{F}^{(i)})\|^2}{\tau_m \sigma^2} \right)
        +
        \frac{2 p_m}{\tau_m \sigma^2 \ln 2 \left( 1 + \frac{p_m \|\mathbf{h}_m(\mathbf{F}^{(i)})\|^2}{\tau_m \sigma^2} \right)}
        \sum_{k=1}^{K} \Re \left\{
        h_{m,k}(\mathbf{f}_k^{(i)})^* (\tilde{\mathbf{h}}_{m,k}^{\prime})^T (\mathbf{f}_k - \mathbf{f}_k^{(i)})
        \right\}
    $}
    \label{eq:lambda_expansion}
    \end{equation}
    \vspace{-10pt} 
    \hrulefill
    \vspace{-10pt} 
\end{figure*}
\vspace{-10pt}

Nevertheless, (P5) retains its non-convex character arising from the unit-modulus restriction in \eqref{eq:P4_norm}. Consequently, we apply a relaxation to $\|f_k\| \le 1$, thereby establishing the convex formulation presented below
\begin{subequations}\label{prob:P6}
\begin{align}
\text{(P6)} \quad \max_{\mathbf{F}} \quad & \Lambda_{r}^{(i+1)}(\mathbf{F}) \label{eq:P6_obj} \\
\text{s.t.} \quad & \cos(\theta_{\text{max}}) \leq \mathbf{f}_k^T \mathbf{e}_3 \leq 1, \quad \forall k, \label{eq:P6_angle} \\
& \|\mathbf{f}_k\| \leq 1, \quad \forall k. \label{eq:P6_norm}
\end{align}
\end{subequations}

Since (P6) constitutes a standard convex optimization formulation, it admits efficient numerical solutions using the CVX toolkit. Furthermore, owing to the relaxation applied to the equality restriction \eqref{eq:P6_norm}, the resulting optimal value of (P6) serves as a theoretical upper bound for the original problem (P5). After obtaining the optimal solutions of the CPU frequencies, we normalize each pointing vector $\mathbf{f}_k$ in $\mathbf{F}$ to satisfy constraint~\eqref{eq:P4_norm}, i.e., $\mathbf{f}_k^{*}=\frac{\mathbf{f}_k}{\|\mathbf{f}_k\|}$.
This normalization only scales $\mathbf{f}_k$ to a unit vector without changing its direction; hence, $\mathbf{f}_k^{*}$ also satisfies~\eqref{eq:P4_angle}. 
\vspace{-10pt}

\subsection{Offloading Optimization}
Given the RA pointing matrix $\mathbf{F}$, the effective channel gains become fixed parameters. Consequently, the remaining optimization for offloading resources $\{\mathbf{f}, \mathbf{p}, \boldsymbol{\tau}\}$ shares the same mathematical formulation. Accordingly, the signal captured by the base station during the $m$-th transmission interval is $\mathbf{y}_m = \sqrt{p_m}\,\mathbf{h}_m(\mathbf{F})\,x_m + \mathbf{n}_m$, where $x_m$ denotes the information symbol satisfying $\mathbb{E}[|x_m|^2]=1$, and $\mathbf{n}_m\sim\mathcal{CN}(\mathbf{0},\sigma^2\mathbf{I})$ characterizes the vector of additive white Gaussian noise. Consequently, problem (2) can be reduced to
\vspace{-7pt}
\begin{subequations}\label{eq:P3_full}
\begin{align}
\text{(P7)}: \max_{\mathbf{y},\,\boldsymbol{\tau},\,\mathbf{f}} \quad & \sum_{m=1}^M \left[ \frac{T f_m}{C} + \tau_m\frac{B}{v_m} L_m(\mathbf{F}^{(i)}) \right] \label{eq:P3_target} \\
\text{s.t.} \quad & y_m + \tau_m p_{c,m} + T r_c f_m^3 \le E_{\max}, \quad \forall m, \label{eq:P3_constr_energy} \\
& \eqref{eq:P1_time}, \eqref{eq:P1_min_bits}, \quad y_m\ge0,\ f_m\ge0,\ \forall m. \label{eq:P3_constr_combined}
\end{align}
\end{subequations}

\noindent
\vspace{-10pt}
Let
\begin{equation}
L_m(\mathbf{F}^{(i)}) = \log_2 \left( 1 + \frac{y_m \|\mathbf{h}_m(\mathbf{F}^{(i)})\|^2}{\tau_m \sigma^2} \right).
\label{eq:Lf_def}
\end{equation}

Upon inspection, the convexity of the constraints \eqref{eq:P3_constr_combined} within (P7) is readily verified. This is because \eqref{eq:P1_min_bits} exhibits the structure of a function lower-bounded by a constant, while the remaining conditions are composed entirely of linear inequalities. In \eqref{eq:P3_constr_energy}, the left-hand side \(y_m + \tau_m p_{c,m} + T r_c f_m^3\) is convex with respected to \((y_m,\tau_m,f_m)\) due to the cubic term with positive coefficient), hence \eqref{eq:P3_constr_energy} is convex. 
For the objective (P7), the term
\(
\tau_m \frac{B}{v_m}\log_2\!\big( 1 + \frac{y_m \|\mathbf{h}_m(\mathbf{F}^{(i)})\|^2}{\tau_m \sigma^2}\big)
\)
is the perspective of the concave function 
\(
\frac{B}{v_m}\log_2\!\big(1+\frac{y_m\|\mathbf{h}_m(\mathbf{F}^{(i)})\|^2}{\sigma^2}\big)
\),
and is therefore jointly concave in \((\tau_m,y_m)\) (perspective preserves concavity \cite{9373363}). The remaining term \(\tfrac{T f_m}{C}\) is linear in \(f_m\). Consequently, (P7) constitutes a convex optimization formulation, which is amenable to efficient resolution via established standard solvers.
\vspace{-10pt}
\subsection{Overall Solution Strategy and Its Complexity}
Algorithm 1 outlines the proposed joint optimization framework. The complexity is $\mathcal{O}(MK + M^{3})$ for the dynamic case and $\tilde{\mathcal{O}}(I_{\text{iter}}(I_{\text{SCA}}K^{3.5} + M^{3}))$ for the static case, where $I_{\text{iter}}$ is the number of outer iterations.

\begin{algorithm}[t]
\caption{Proposed SAHO Algorithm for (P3)}
\label{alg:SAHO}
\begin{algorithmic}[1]
\STATE \textbf{Initialization:} Set iteration index $i=0$. Initialize feasible $\mathbf{F}^{(0)}$ and offloading parameters $\{ \mathbf{f}^{(0)}, \mathbf{p}^{(0)}, \bm{\tau}^{(0)} \}$.
\IF{\textit{Dynamic rotating scenario}}
    \STATE Compute optimal pointing matrix $\mathbf{F}^{*}$ directly utilizing the closed-form solution in (\ref{eq:f_opt}).
    \STATE Update offloading resources $\{\mathbf{f}^{*}, \mathbf{p}^{*}, \bm{\tau}^{*}\}$ by solving problem (P3) with fixed $\mathbf{F}^{*}$.
\ELSIF{\textit{Static rotating scenario}}
    \REPEAT
        \STATE Update $\mathbf{F}^{(i+1)}$ by solving the SCA approximation problem (P6) with fixed offloading parameters $\{\mathbf{f}^{(i)}, \mathbf{p}^{(i)}, \bm{\tau}^{(i)}\}$.
        \STATE Update effective channel gains $\mathbf{h}_m(\mathbf{F}^{(i+1)})$.
        \STATE Update offloading resources $\{\mathbf{f}^{(i+1)}, \mathbf{p}^{(i+1)}, \bm{\tau}^{(i+1)}\}$ by solving problem (P3) with fixed $\mathbf{F}^{(i+1)}$.
        \STATE Update iteration index $i \leftarrow i + 1$.
    \UNTIL{The weighted sum computation rate converges.}
    \STATE Set $\{\mathbf{F}^*, \mathbf{f}^*, \bm{\tau}^*, \mathbf{p}^*\} \leftarrow \{\mathbf{F}^{(i)}, \mathbf{f}^{(i)}, \bm{\tau}^{(i)}, \mathbf{p}^{(i)}\}$.
\ENDIF
\STATE \textbf{Output:} Optimal solution $\{\mathbf{F}^*, \mathbf{f}^*, \bm{\tau}^*, \mathbf{p}^*\}$.
\end{algorithmic}
\end{algorithm}
\vspace{-10pt}


\begin{figure*}[t]
    \centering
    \subfloat[Total computation rate versus iteration number for different directivity factors $p$.]{
        \includegraphics[width=0.25\textwidth]{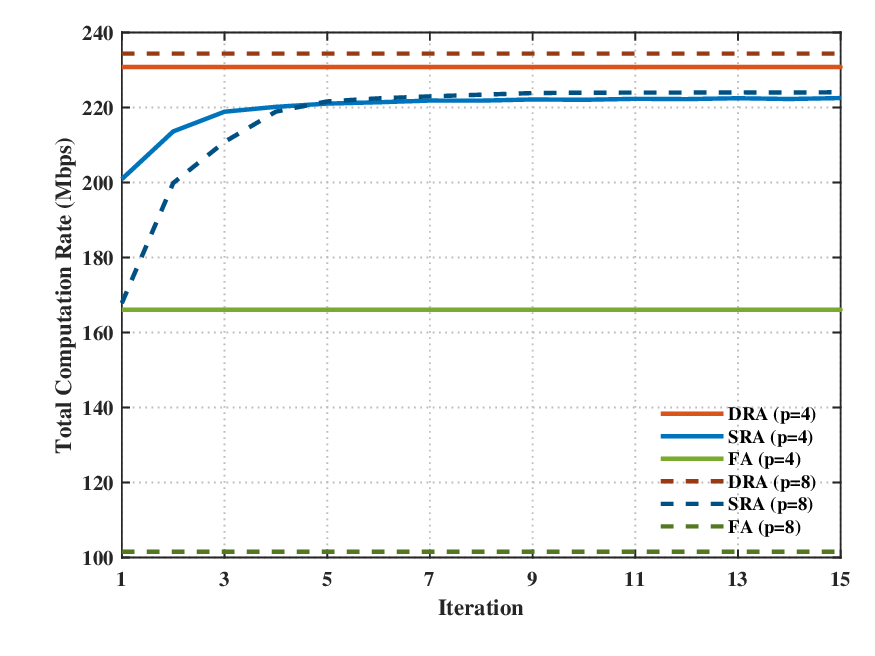} 
        \label{fig:sub_convergence}
    }
    \hfill 
    \subfloat[Total computation rate versus maximum rotation angle $\theta_{\max}$.]{
        \includegraphics[width=0.25\textwidth]{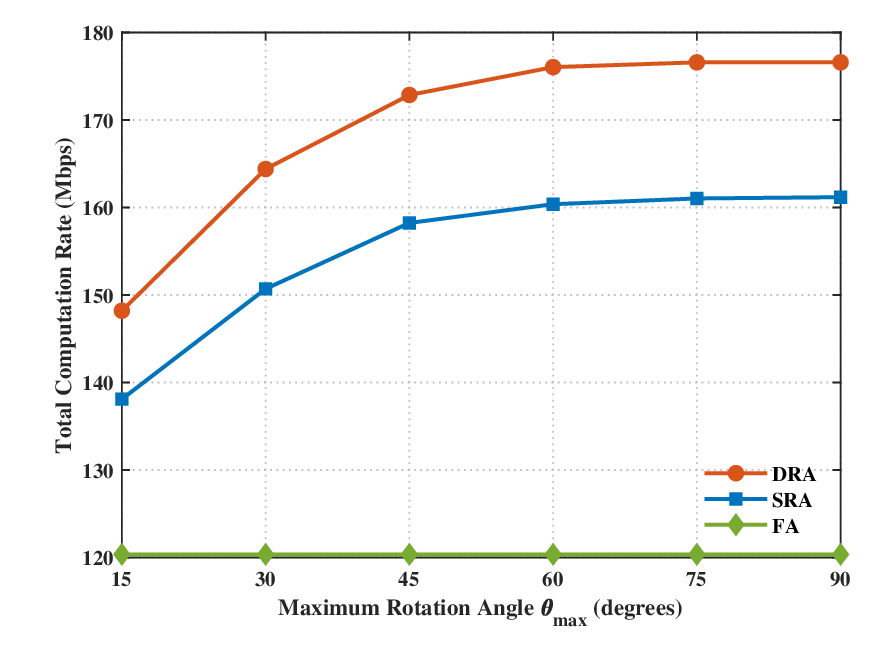} 
        \label{fig:sub_theta}
    }
    \hfill 
    \subfloat[Total computation rate versus number of antennas $K$.]{
        \includegraphics[width=0.25\textwidth]{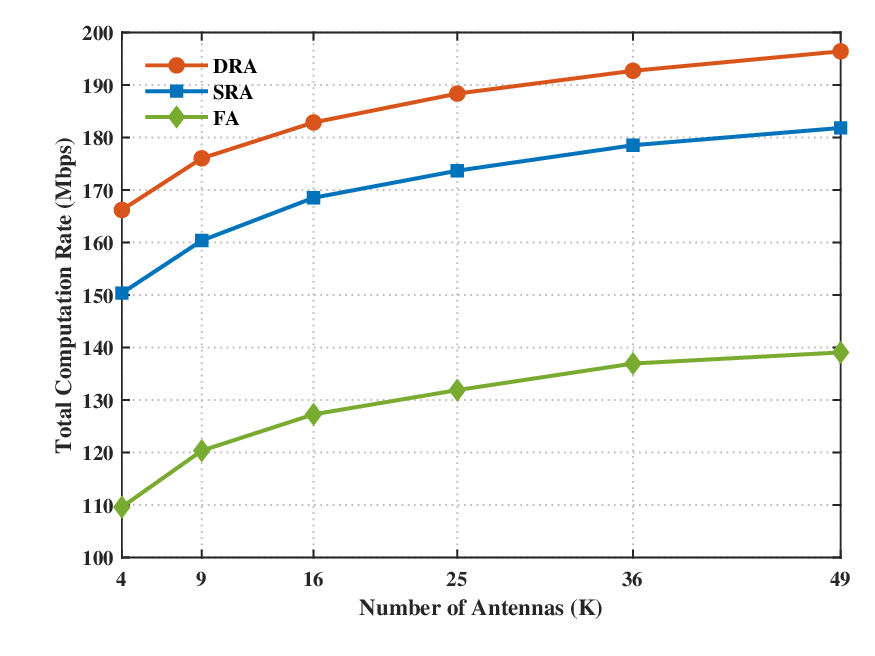} 
        \label{fig:sub_antennas}
    }
    
    \caption{Performance comparison. (a) Convergence behavior; (b) Impact of rotation angle limit; (c) Impact of antenna number.}
    \label{fig:overall_performance}
\end{figure*}

\section{Simulation Results}
\label{sec:simulation}

This section is dedicated to assessing the efficacy of the developed RA-assisted MEC framework. We assume a carrier frequency of 2.4 GHz ($\lambda = 0.125$ m). The BS is equipped with a UPA of $K = K_x \times K_y$ RAs with spacing $\Delta = \lambda/2$. Unless otherwise stated, the default settings are: $K = 9$ ($3\times3$), $p = 4$, $\theta_{\max} = \pi/3$, $M = 4$ users, $T = 1$ s, $B = 10$ MHz, $C = 1000$ cycles/bit, $E_{\max} = 10$ J, $\sigma^2 = -100$ dBm, and Rician factor $\kappa_m=1$. User terminals are spatially dispersed with uniform density within a 3D cylindrical volume enclosing the central BS. Specifically, the horizontal distance of each user from the BS follows a uniform distribution in $[20, 50]$ m, and the user height is uniformly distributed in $[10, 30]$~m. 

We benchmark the proposed Static Rotatable Antenna (SRA) scheme against two baselines: a Fixed Antenna (FA) system ($f=[0,0,1]^T$) serving as a lower bound, and a theoretical Dynamic Rotatable Antenna (DRA) scheme where antennas reorient per time slot, serving as an upper bound.

Fig. 2(a) confirms the efficiency of our algorithm, which converges within 10 iterations. More critically, it reveals a fundamental limitation of fixed antennas: as directivity $p$ increases, the FA performance collapses due to severe beam misalignment. In contrast, the RA scheme exploits mechanical steering to maintain alignment, doubling the computation rate at $p=8$. 

Fig. 2(b) reveals that the computation rate improves with $\theta_{\max}$ but saturates beyond $60^{\circ}$. This suggests that a restricted rotation range of $\pm60^{\circ}$ covers the majority of user distributions. Crucially, this implies that low-complexity, limited-range rotators are sufficient to achieve near-optimal performance, avoiding the hardware costs of full-steering mechanisms while still outperforming fixed antennas.

Finally, Fig. 2(c) examines the impact of the antenna number, $K$. While expanding the array size generally improves the computation rate, the marginal gain diminishes as the system becomes energy-limited. Throughout this range, the RA scheme consistently outperforms the fixed baseline, confirming its ability to correct beam misalignment and fully unlock the potential array gain.
\vspace{-7pt}

\section{Conclusion}
This paper investigated an RA-assisted MEC system to mitigate beam misalignment through mechanical antenna rotation. We formulated a weighted sum computation rate maximization problem and proposed a SAHO algorithm to jointly optimize RA orientation and offloading resources. Specifically, we proposed a sequential solution based on closed-form antenna pointing for dynamic scenarios to show the maximum potential of RA, and employed an SCA-based AO scheme for static scenarios to effectively handle coupled constraints. Simulation results confirm that the proposed design significantly outperforms fixed-antenna benchmarks, particularly under high antenna directivity, validating its practical effectiveness.
\vspace{-7pt}

\bibliographystyle{IEEEtran}

\bibliography{refs}

\end{document}